\documentclass[aps,prd,10pt,showpacs,showkeys,preprintnumbers,superscriptaddress,nobibnotes,floatfix,longbibliography,nofootinbib]{revtex4-2}

\usepackage{amsmath,amsfonts,amssymb,mathrsfs,graphicx,color,longtable}
\usepackage{hyperref}
\usepackage{bm}
\usepackage{array}
\usepackage{color}
\usepackage{enumitem}
\usepackage[explicit]{titlesec}
\usepackage{float}  
\usepackage{subfigure}
\usepackage{multirow,bigdelim}
\usepackage{blindtext}
\usepackage{rotating}
\usepackage{stfloats}
\usepackage{ulem}
\usepackage[ddmmyy,24hr]{datetime}

\def\U{{}^{235}\text{U}}
\def\Um{{}^{238}\text{U}}
\def\Pu{{}^{239}\text{Pu}}
\def\Pum{{}^{241}\text{Pu}}

\def\sigmacomb{\sigma_{\text{Pu}}}
\def\ibdunit{\times\,10^{-43}\,\rm{cm^2/fission}}
\def\ibdunits{\left[10^{-43}\,\rm{cm^2/fission} \right]}

\newcolumntype{C}[1]{>{\centering\let\newline\\\arraybackslash\hspace{4pt}}m{#1}}

\graphicspath{{figures/}}

\begin{document}

\title{Model-Independent Determination of Isotopic Cross Sections per Fission for Reactor Antineutrinos}

\author{Yu-Feng Li}
\email{liyufeng@ihep.ac.cn}
\affiliation{Institute of High Energy Physics,
Chinese Academy of Sciences, Beijing 100049, China}
\affiliation{School of Physical Sciences, University of Chinese Academy of Sciences, Beijing 100049, China}

\author{Zhao Xin}
\email{xinzhao@ihep.ac.cn (corresponding author)}
\affiliation{Institute of High Energy Physics,
Chinese Academy of Sciences, Beijing 100049, China}
\affiliation{School of Physical Sciences, University of Chinese Academy of Sciences, Beijing 100049, China}

\date{\dayofweekname{\day}{\month}{\year} \ddmmyydate\today, \currenttime}

\begin{abstract}

Model-independent reactor isotopic cross sections per fission are determined by global fits of the reactor antineutrino data from High-Enriched Uranium (HEU) reactor rates, Low-Enriched Uranium (LEU) reactor rates, and reactor fuel evolution data. Taking account of the implicit quasi-linear relationship between the fission fractions of $\Pu$ and $\Pum$ in the LEU reactor data, the Inverse-Beta-Decay (IBD) yields and their correlations of the fissionable isotopes $\U$, $\Um$, and Pu's are obtained. The data-driven isotopic IBD yields provide an anomaly-free model for the reactor isotopic cross sections per fission, where better than 1\% accuracy of the expected reactor IBD yields can be achieved for future experiments.

\end{abstract}

\maketitle


\section{Introduction}

Reactor antineutrinos originate from the beta decays of the neutron-rich fission fragments produced by the fission of four main isotopes $\U$, $\Um$, $\Pu$, and $\Pum$~\cite{Bemporad:2001qy,Huber:2016fkt,Hayes:2016qnu}.
The nuclear reactors are one of the most important sources of electron antineutrinos, which are applied to the study of the fundamental properties of neutrinos~\cite{ParticleDataGroup:2020ssz}, from the first discovery of the neutrino~\cite{Cowan:1956rrn} to the studies of the neutrino oscillation phenomenon~\cite{KamLAND:2002uet,DayaBay:2012fng,RENO:2012mkc,DoubleChooz:2011ymz} and new physics beyond the Standard Model (SM)~\cite{Kopp:2013vaa,Gariazzo:2015rra}. 

The theoretical fluxes of reactor antineutrinos are crucial inputs for reactor experiments, which can be obtained with two complementary methods~\cite{Huber:2016fkt,Hayes:2016qnu}. One is usually called the conversion method, in which the antineutrino fluxes are converted from the measured cumulative beta spectra, such as those of $\U$, $\Pu$, and $\Pum$ at Institut Laue-Langevin (ILL)~\cite{VonFeilitzsch:1982jw,Schreckenbach:1985ep,Hahn:1989zr}  or at Kurchatov Institute (KI)~\cite{Kopeikin:2021ugh}  from the thermal-neutron induced fission process,
and that of $\Um$ at FRM~\uppercase\expandafter{\romannumeral2} in Garching~\cite{Haag:2013raa} from the fission process induced by fast neutrons.
The other method is the \textit{`ab initio'} summation method, which directly sums the energy spectra of all beta decay branches from all the fission fragments. These two methods have motivated us to construct different theoretical models for reactor antineutrino fluxes.
The Huber-Mueller (HM) model is arguably the most well-known reactor flux model, in which the antineutrino fluxes of $\U$, $\Pu$ and $\Pum$ are obtained by Huber~\cite{Huber:2011wv} with the conversion method,
and the flux of $\Um$ is obtained by Mueller~\textit{et al.}~\cite{Mueller:2011nm} using the summation method. Hayen, Kostensalo, Severijns, and Suhonen~\cite{Hayen:2019eop} improved the conversion method by including the forbidden transitions, and proposed a correction to the HM model, which is abbreviated as the HKSS model. In Ref.~\cite{Kopeikin:2021ugh}, a new measurement of the cumulative beta spectra of $\U$ is considered and a corrected $\U$ antineutrino flux is obtained. Moreover, in the KI model, a converted $\Um$ antineutrino flux is also presented based on the measured beta spectra of $\Um$ in FRM~\uppercase\expandafter{\romannumeral2}. On the other hand, with the development of the nuclear database and the Pandemonium free data~\cite{GREENWOOD1992514,Tengblad:1989db,Algora:2010zz,IGISOL:2015ifm,Valencia:2016rlr,Rice:2017kfj,Guadilla:2019aiq}, the updated summation model by Estienne, Fallot \textit{et al.}~\cite{Estienne:2019ujo} is also presented (EF model). More details on the properties and comparisons of these models are summarized in Ref.~\cite{Giunti:2021kab}. In addition, some other studies on the theoretical reactor antineutrino fluxes~\cite{Schreckenbach:1985ep,Vogel:2007du,Silaeva:2020msh,Fallot:2012jv,Li:2019quv,Fang:2020emq} are also established to offer precise predictions of the experimental observation.  

However, the experimental measurements of reactor antineutrino rates and energy spectra have
shown anomalous results compared to the theoretical predictions.
One is a $6\%$ deficit between the experimental and theoretical reactor antineutrino rates~\cite{Mention:2011rk} (i.e., the rate anomaly). The other is a spectral excess between the experimental and theoretical reactor antineutrino spectra~\cite{DayaBay:2015lja,NEOS:2016wee,DoubleChooz:2014kuw} (i.e., the shape anomaly).
These anomalies have questioned the foundation of all the aforementioned reactor antineutrino flux models. In Ref.~\cite{Giunti:2021kab}, it is shown that both the EF summation model and the KI conversion model tend to reduce the $\U$ flux, and provide a plausible solution to the reactor rate anomaly, but the solution to the reactor shape anomaly is still elusive. Therefore, it would be important to obtain a model-independent determination of the reactor antineutrino fluxes, which is the main focus of this work.
By using the quasi-linear relationship between the fission fractions of $\Pu$ and $\Pum$, we present a method to directly obtain the isotopic Inverse-Beta-Decay (IBD) yields from global fits of the reactor antineutrino data from High-Enriched Uranium (HEU) reactor rates, Low-Enriched Uranium (LEU) reactor rates, and reactor fuel evolution data.
The IBD yields and their correlations of the fissionable isotopes $\U$, $\Um$, and Pu's are obtained, and can be applied to future reactor experiments. Taking the representative values of fission fractions for typical commercial reactors, the accuracy of the expected IBD yield can be better than 1\%.


This work is organized as follows. In Section~\ref{sec.method}, we discuss how we deduce the data-driven isotopic IBD yields. In Section~\ref{sec.results}, we provide results of the global fits of the reactor data. We present and discuss different fitting results of the HEU reactor rates, fuel evolution data and LEU reactor rates from Section~\ref{subsec.HEU} to Section~\ref{subsec.evo_rates}. Moreover, the application how to apply the data-driven isotopic IBD yields to the prediction of future reactor experiments is illustrated in Section~\ref{subsec.app}. Finally, the concluding remarks are summarized in Section~\ref{sec.conclusion}.


\section{Model independent method}
\label{sec.method}

The reactor IBD yield $\sigma_{f,a}$ is used to illustrate the measured reactor antineutrino rates known as cross section per fission as well:
\begin{align}
    \sigma_{f,a}=\sum_i f_i^a \sigma_i\,,
\end{align}
where $a$ represents the sequence of the experiments, $\sigma_i$ is the isotopic IBD yield for a certain fissionable isotope ($i=$235, 238, 239 and 241 for $\U$, $\Um$, $\Pu$ and $\Pum$, respectively), and $f_i^a$ is the fission fraction of the isotope $i$ in the experiment $a$. For a certain fissionable isotope $i$, whose isotopic IBD yield is defined as
\begin{align}
    \sigma_i=\int_{E_\nu^{\text{thr}}}^{E_\nu^{\text{max}}} \Phi_i(E_\nu)\sigma_{\text{IBD}}(E_\nu) d E_\nu,
\end{align}
where $E_\nu$ is the antineutrino energy, $\Phi(E_\nu)$ is the antineutrino flux and $\sigma_{\text{IBD}}(E_\nu)$ is the IBD cross section~\cite{Strumia:2003zx,Kurylov:2002vj}. The theoretical isotopic IBD yield for a certain isotope $i$ from different flux models~\cite{Mueller:2011nm,Haag:2013raa,Huber:2011wv,Estienne:2019ujo,Hayen:2019eop,Kopeikin:2021ugh} has been revisited in~\cite{Giunti:2021kab}, and will be used in the current analysis. 


The nuclear reactors of reactor antineutrino experiments can be divided into two categories: the commercial reactors mainly composited with $\U\, (\sim 50-60\%)$ and $\Pu\, (\sim 25-35\%)$, which is also known as the LEU nuclear power reactors; and the research reactors with practically pure $\U$ known as HEU research reactors. Therefore, the reactor antineutrino data are dominated by $\U$ and $\Pu$ and are provided in Table~\ref{tab.rates_2020}, which is reordered from Table \uppercase\expandafter{\romannumeral4} of Ref.~\cite{Giunti:2021kab} by considering different types of reactors. 
In Table~\ref{tab.rates_2020}, $a$ is the index of each experiment,
$f^{a}_{i}$'s
are the effective fission fractions of the four isotopes,
$\sigma_{f,a}^{\text{exp}}$
is the experimental IBD yield
in units of $10^{-43} \text{cm}^{2}/\text{fission}$,
$\delta_{a}^{\text{exp}}$ is the total relative experimental statistical plus systematic uncertainty,
$\delta_{a}^{\text{cor}}$ is the part of the relative experimental systematic uncertainty
which is correlated in each group of experiments indicated by the braces;
$L_{a}$ is the source-detector distance. In addition, we also consider the fuel evolution data from Daya Bay~\cite{DayaBay:2017jkb} and RENO~\cite{RENO:2018pwo}, each of which provided the experimental IBD yields as a function of the fission fractions. 


\begin{table*}[ht]
\centering
\begin{tabular}{c|cccccccccc}
\hline
\hline
rate
&
$a$
&
Experiment
&
$f^{a}_{235}$
&
$f^{a}_{238}$
&
$f^{a}_{239}$
&
$f^{a}_{241}$
&
$\sigma_{f,a}^{\text{exp}} $
&
$\delta_{a}^{\text{exp}}$ [\%]
&
$\delta_{a}^{\text{cor}}$ [\%]
&
$L_{a}$ [m]
\\
\hline
\multirow{18}*{LEU} & 1	&Chooz 			 &0.496	&0.087	&0.351	&0.066	&6.12 &3.2	&0	 &$\approx 1000$\\
~ & 2 	&Palo Verde      &0.600	&0.070	&0.270	&0.060	&6.25 &5.4	&0	 &$\approx 800$\\
~ & 3 	&Daya Bay		 &0.564	&0.076	&0.304	&0.056	&5.94 &1.5	&0	 &$\approx 550$\\
~ & 4 	&RENO 			 &0.571	&0.073	&0.300	&0.056	&5.85 &2.1	&0	 &$\approx 411$\\
~ & 5 	&Double Chooz 	 &0.520	&0.087	&0.333	&0.060	&5.71 &1.1	&0	 &$\approx 415$\\
~ & 6 	&Bugey-4 		 &0.538	&0.078	&0.328	&0.056	&5.75  &1.4	&\rdelim\}{2}{20pt}[1.4]	 &$15$\\
~ & 7 	&Rovno91 		 &0.614	&0.074	&0.274	&0.038	&5.85	 &2.8	&                      	&$18$\\
~ & 8 	&Rovno88-1I 		 &0.607	&0.074	&0.277	&0.042	&5.70	&6.4	&\rdelim\}{2}{20pt}[3.1] \rdelim\}{5}{20pt}[2.2]	 &$18$\\
~ & 9 	&Rovno88-2I 		 &0.603	&0.076	&0.276	&0.045	&5.89	 &6.4	&                                               	&$17.96$\\
~ & 10 	&Rovno88-1S 		 &0.606	&0.074	&0.277	&0.043	&6.04	 &7.3	&\rdelim\}{3}{45pt}[3.1]                        	&$18.15$\\
~ & 11 	&Rovno88-2S 		 &0.557	&0.076	&0.313	&0.054	&5.96	 &7.3	&                                              	&$25.17$\\
~ & 12 	&Rovno88-3S 		 &0.606	&0.074	&0.274	&0.046	&5.83	 &6.8	&                                               	&$18.18$\\
~ & 13 	&Bugey-3-15 		 &0.538	&0.078	&0.328	&0.056	&5.77  &4.2	&\rdelim\}{3}{20pt}[4.0]                       	&$15$\\
~ & 14 	&Bugey-3-40 		 &0.538	&0.078	&0.328	&0.056	&5.81	 &4.3	&                                              	&$40$\\
~ & 15 	&Bugey-3-95 		 &0.538	&0.078	&0.328	&0.056	&5.35	 &15.2	&                                               	&$95$\\
~ & 16 	&Gosgen-38 		 &0.619	&0.067	&0.272	&0.042	&5.99 &5.4	&\rdelim\}{3}{20pt}[2.0] \rdelim\}{4}{20pt}[3.8]  	&$37.9$\\
~ & 17 	&Gosgen-46 		 &0.584	&0.068	&0.298	&0.050	&6.09	 &5.4	&     	&$45.9$\\
~ & 18 	&Gosgen-65 		 &0.543	&0.070	&0.329	&0.058	&5.62 &6.7	&                                                &$64.7$\\
\hline
\multirow{9}*{HEU} & 19 	&ILL 			 &1	&0	&0	&0	&5.30	&9.1	&                                                	&$8.76$\\
~ & 20 	&Krasnoyarsk87-33 	 &1	&0	&0	&0	&6.20	 &5.2	&\rdelim\}{2}{20pt}[4.1]	 &$32.8$\\
~ & 21 	&Krasnoyarsk87-92 	 &1	&0	&0	&0	&6.30	 &20.5	&                        &$92.3$\\
~ & 22 	&Krasnoyarsk94-57 	 &1	&0	&0	&0	&6.26	 &4.2	&0                      	 &$57$\\
~ & 23 	&Krasnoyarsk99-34 	 &1	&0	&0	&0	&6.39	 &3.0	&0                      	 &$34$\\
~ & 24 	&SRP-18 		 &1	&0	&0	&0	&6.29 &2.8	&0	 &$18.2$\\
~ & 25 	&SRP-24 		 &1	&0	&0	&0	&6.73 &2.9	&0	 &$23.8$\\
~ & 26 	&STEREO		 &1 &0 &0 &0	&6.34  &2.5  &0	  &$9-11$\\
\hline
\hline
\end{tabular}
\caption{\label{tab.rates_2020}
List of the experiments which measured the absolute reactor antineutrino fluxes.
For each experiment numbered with the index $a$:
$f^{a}_{235}$,
$f^{a}_{238}$,
$f^{a}_{239}$, and
$f^{a}_{241}$
are the effective fission fractions of the four isotopes
$^{235}\text{U}$,
$^{238}\text{U}$,
$^{239}\text{Pu}$, and
$^{241}\text{Pu}$,
respectively;
$\sigma_{f,a}^{\text{exp}}$
is the experimental IBD yield
in units of $10^{-43} \text{cm}^{2}/\text{fission}$;
$\delta_{a}^{\text{exp}}$ is the total relative experimental statistical plus systematic uncertainty,
$\delta_{a}^{\text{cor}}$ is the part of the relative experimental systematic uncertainty
which is correlated in each group of experiments indicated by the braces;
$L_{a}$ is the source-detector distance. 
}
\end{table*}

\begin{figure}[ht]
    \centering
    \subfigure[LEU reactor rates]{\label{fig.LEU_deltaf}
 	\includegraphics[width=0.48\textwidth]{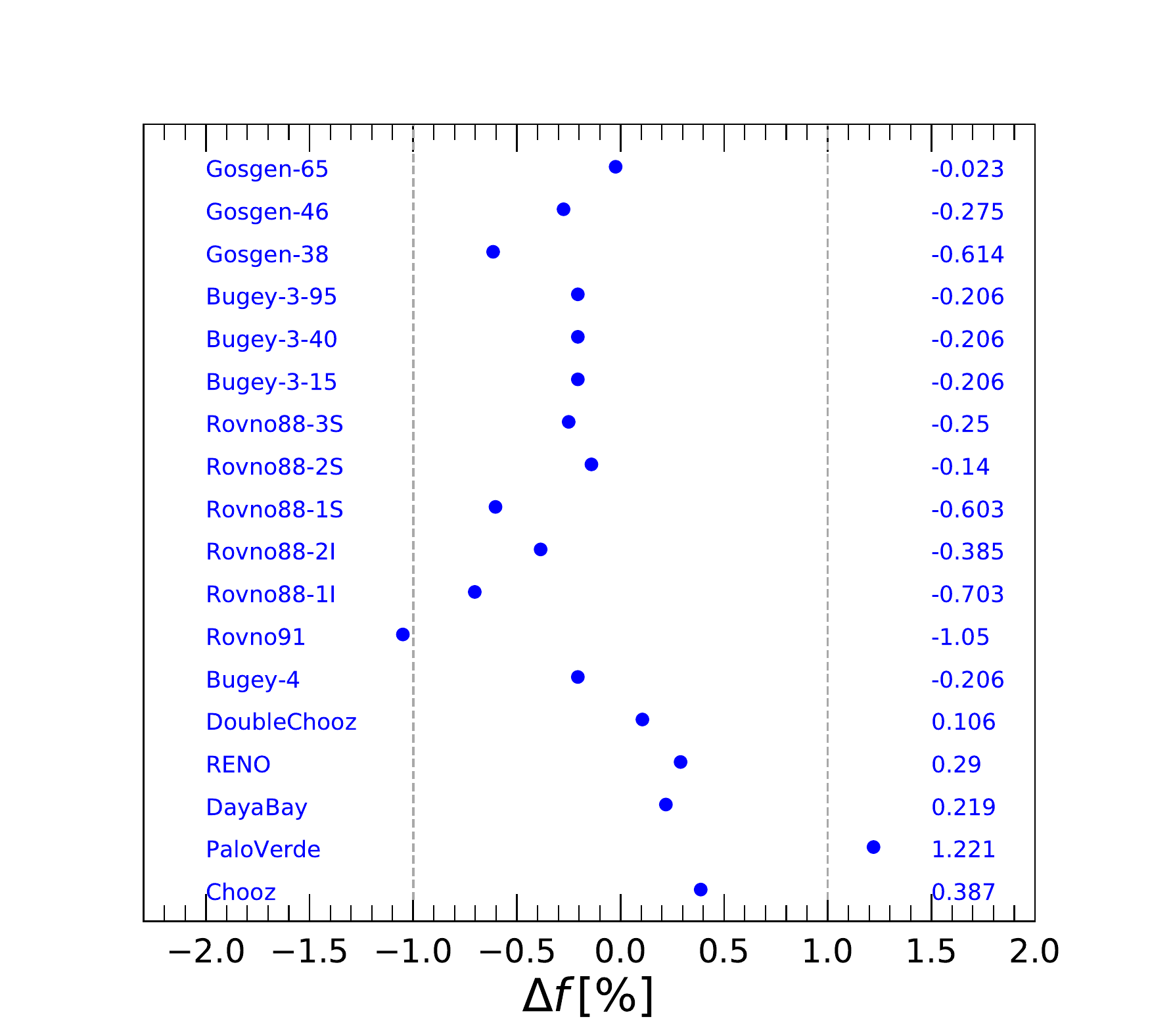}}
 	\subfigure[Fuel evolution data]{\label{fig.evo_deltaf}
  	\includegraphics[width=0.48\textwidth]{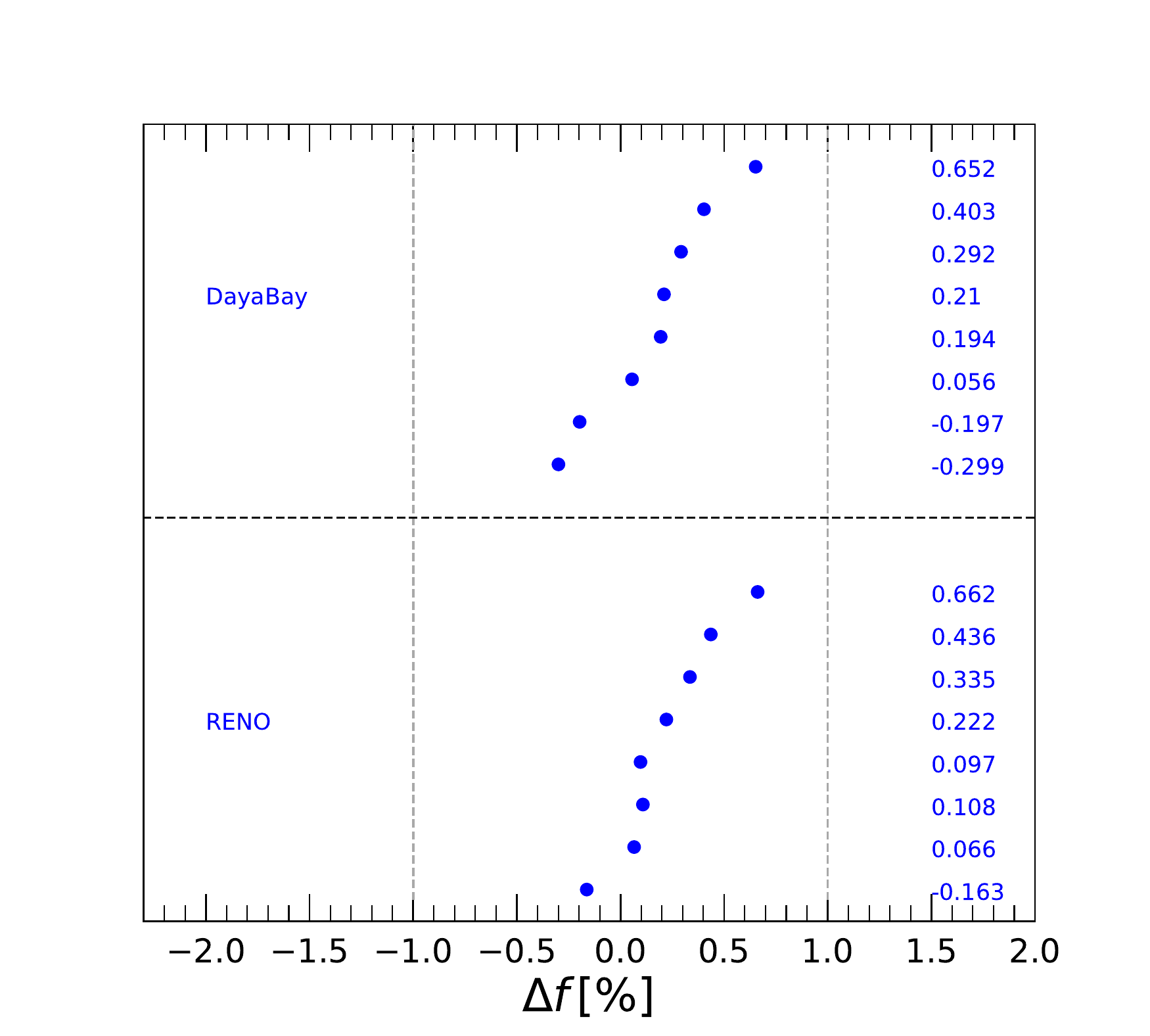}}
	\label{fig.deltafs}
    \caption{Residual fission fractions of $\Pum$ for the LEU rate data evaluated from Table~\ref{tab.rates_2020} and fuel evolution data from Daya Bay\cite{DayaBay:2017jkb} and RENO~\cite{RENO:2018pwo} . Most of data points are smaller than $1\%$, while only two points (Rovon 91 and Palo Verde) are slightly larger than $1\%$.}
\end{figure}

Based on the intrinsic reactor fuel burnup relation, it is widely conceived that there is a quasi-liner relationship~~\cite{DayaBay:2021dqj} between the fission fractions of $\Pum$ and $\Pu$ in LEU reactors:
\begin{align}
    f_{241}=k\cdot f_{239}\,, 
    \label{eq.k}
\end{align}
where $f_{239}$ and $f_{241}$ are the fission fractions of $\Pu$ and $\Pum$, respectively, and $k$ is the fitting linear coefficient from reactor rates or evolution data, which are shown in Table~\ref{tab.k} for different groups of data sets.
In this work, we always use the value from the combined data with $k=0.177$.
Therefore, the IBD yield for a certain LEU reactor antineutrino experiment can be expressed as
\begin{align}
    \sigma_{f,a} = & f_{235}^a\cdot \sigma_{235} + f_{238}^a\cdot \sigma_{238} +f_{239}^a\cdot \sigma_{239} 
     + f_{241}^a\cdot \sigma_{241}\nonumber\\
    = & f_{235}^a\cdot \sigma_{235}+f_{238}^a\cdot \sigma_{238}+f_{239}^a\cdot \sigmacomb
     + (f_{241}^a-k\cdot f_{239}^a)\cdot \sigma_{241}\nonumber\\
    = & f_{235}^a\cdot \sigma_{235}+f_{238}^a\cdot \sigma_{238}+f_{239}^a\cdot \sigmacomb
     + \Delta f^a \cdot \sigma_{241} , 
    \label{eq.sigma_fa}
\end{align}
where $a$ represents the index of the experiment, and
\begin{align}
\sigmacomb = \sigma_{239}+k\cdot \sigma_{241}\,,
\end{align}
is the combined isotopic IBD yield of Pu's, with $k$ describing the linear relationship between $\Pu$ and $\Pum$. 
$\Delta f^a = f_{241}^a-k\cdot f_{239}^a$ is the residual fission fraction of $\Pum$, which are shown in Figure~\ref{fig.LEU_deltaf} for LEU reactor rates (left panel) and Figure~\ref{fig.evo_deltaf} (right panel) for fuel evolution data. From the figure, one can note that most of the residual fission fractions of $\Pum$ are smaller than $1\%$, which indicates that model dependence on the IBD yield corrections of the residual $\Pum$ should be negligible.


\begin{table}[ht]
    \centering
    \setlength{\tabcolsep}{11mm}{
    \begin{tabular}{|c|c|}
    \hline
    \hline
         Data sets & $k$ \\
         \hline
         LEU reactors rates & \quad $0.173$ \quad \\
         Daya Bay & \quad $0.183$ \quad \\
         RENO & \quad $0.185$ \quad \\
         Evolution data & \quad $0.184$ \quad \\
         Combined data & \quad $0.177$ \quad \\
             \hline
    \end{tabular}}
    \caption{The fitting coefficients $k$'s obtained from different groups of reactor data. In our analysis, we always use the value from the combined data with $k=0.177$.}
    \label{tab.k}
\end{table}




The least-squares function based on Wilks' theorem~\cite{Wilks:1938dza} adopted in this work can be written as follows:
\begin{align}
    \chi^2 = \sum_{a,b} \left(\sigma_{f,a}^{\text{exp}}-\sigma_{f,a}^{\text{fit}}\right)(V^{\text{exp}})^{-1}_{ab}\left(\sigma_{f,b}^{\text{exp}}-\sigma_{f,b}^{\text{fit}}\right),
    \label{eq.chi2_3d}
\end{align}
where $(a,b)$ are the indexes of the reactor rates or fuel evolution data, $V^{\text{exp}}$ is the experimental covariance matrix, $\sigma_{f,a}^{\text{exp}}$ is the experimental IBD yield, and according to Eq.~(\ref{eq.sigma_fa}), $\sigma_{f,a}^{\text{fit}}$ is the predicted IBD yield to be fitted in the analysis and can be expressed as
\begin{align}
    \sigma_{f,a}^{\text{fit}}=   f_{235}^a\cdot\sigma_{235}^{\text{fit}} + f_{238}^a\cdot\sigma_{238}^{\text{fit}} 
     + f_{239}^a\cdot \sigmacomb^{\text{fit}} + \Delta f^a\cdot \sigma_{241}^{\text{mod}},
    \label{eq.sigma_fa_k}
\end{align}
where $\sigma_{235}^{\text{fit}}$, $\sigma_{238}^{\text{fit}}$ and $\sigma_{\text{Pu}}^{\text{fit}}$ are the physical parameters which represent the best-fit isotopic IBD yields for $\U$, $\Um$, and Pu's, respectively. Meanwhile, $\Delta f^a \cdot \sigma_{241}^{\rm{mod}}$ is the model correction for the residual $\Pum$. 




\section{Fits of reactor experimental data}
\label{sec.results}

\begin{figure*}
    \centering
    \includegraphics[width=1.05\textwidth]{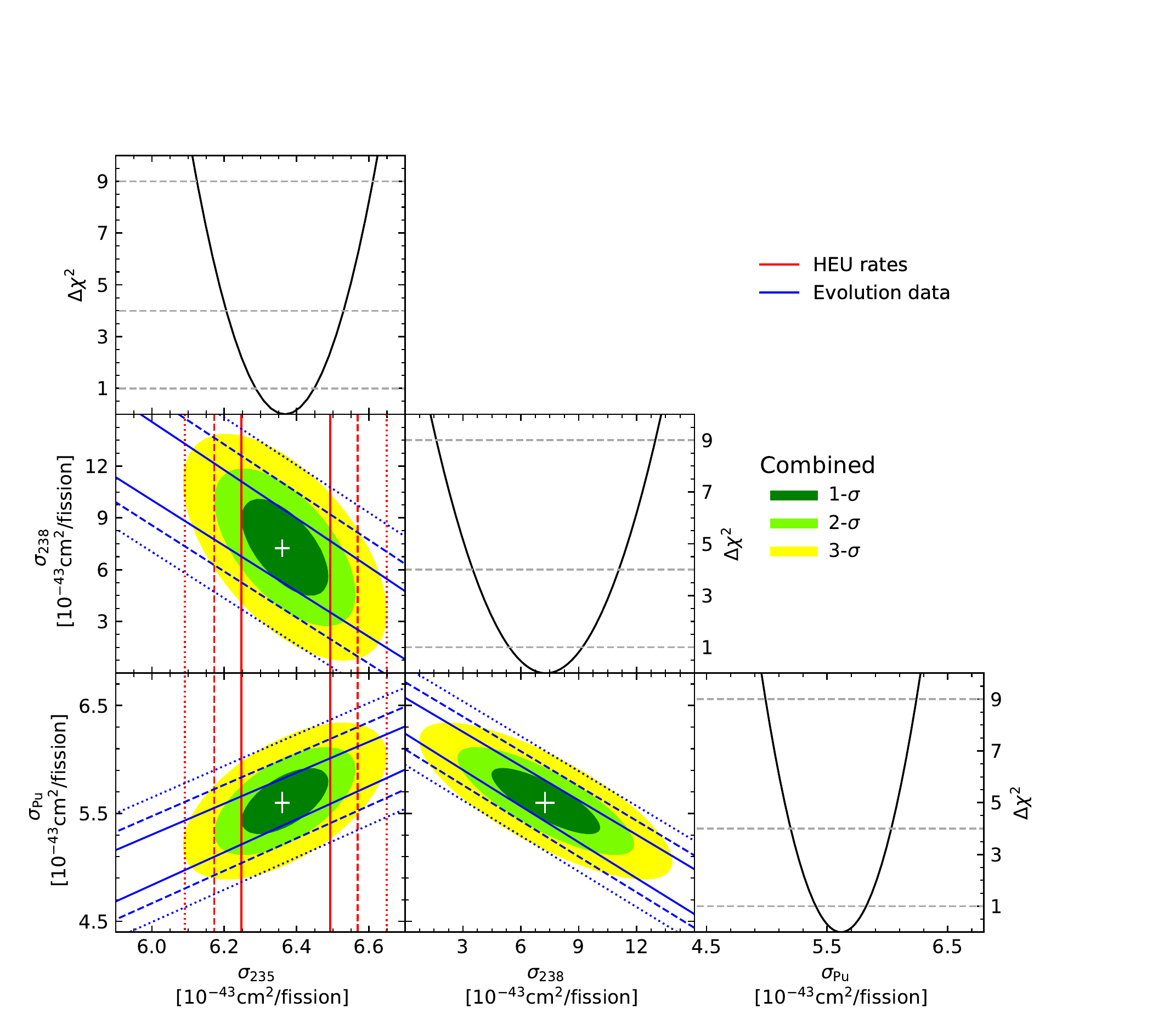}
    \caption{One-dimensional and two-dimensional marginalized distribution with $1\sigma$, $2\sigma$ and $3\sigma$ contours of $\sigma_{235}$, $\sigma_{238}$ and $\sigmacomb$ provided by a fit of \textbf{evolution data + HEU reactor rates}. The red solid, dashed and dotted contours show the allowed regions of \textbf{HEU reactor rates} at $1\sigma$, $2\sigma$ and $3\sigma$. The blue solid, dashed and dotted contours show the allowed regions of \textbf{evolution data} at $1\sigma$, $2\sigma$ and $3\sigma$. The darkgreen, lawngreen and yellow shaded regions show the results of the allowed regions of \textbf{evolution data + HEU reactor rates}. The three black curves show one-dimensional $\Delta \chi^2$ profiles for $\sigma_{235}$, $\sigma_{238}$ and $\sigma_{\rm{Pu}}$ fitted with \textbf{evolution data + HEU reactor rates}. }
    \label{fig.DYB_RENO_HEU}
\end{figure*}

\begin{figure*}
    \centering
    \includegraphics[width=1.05\textwidth]{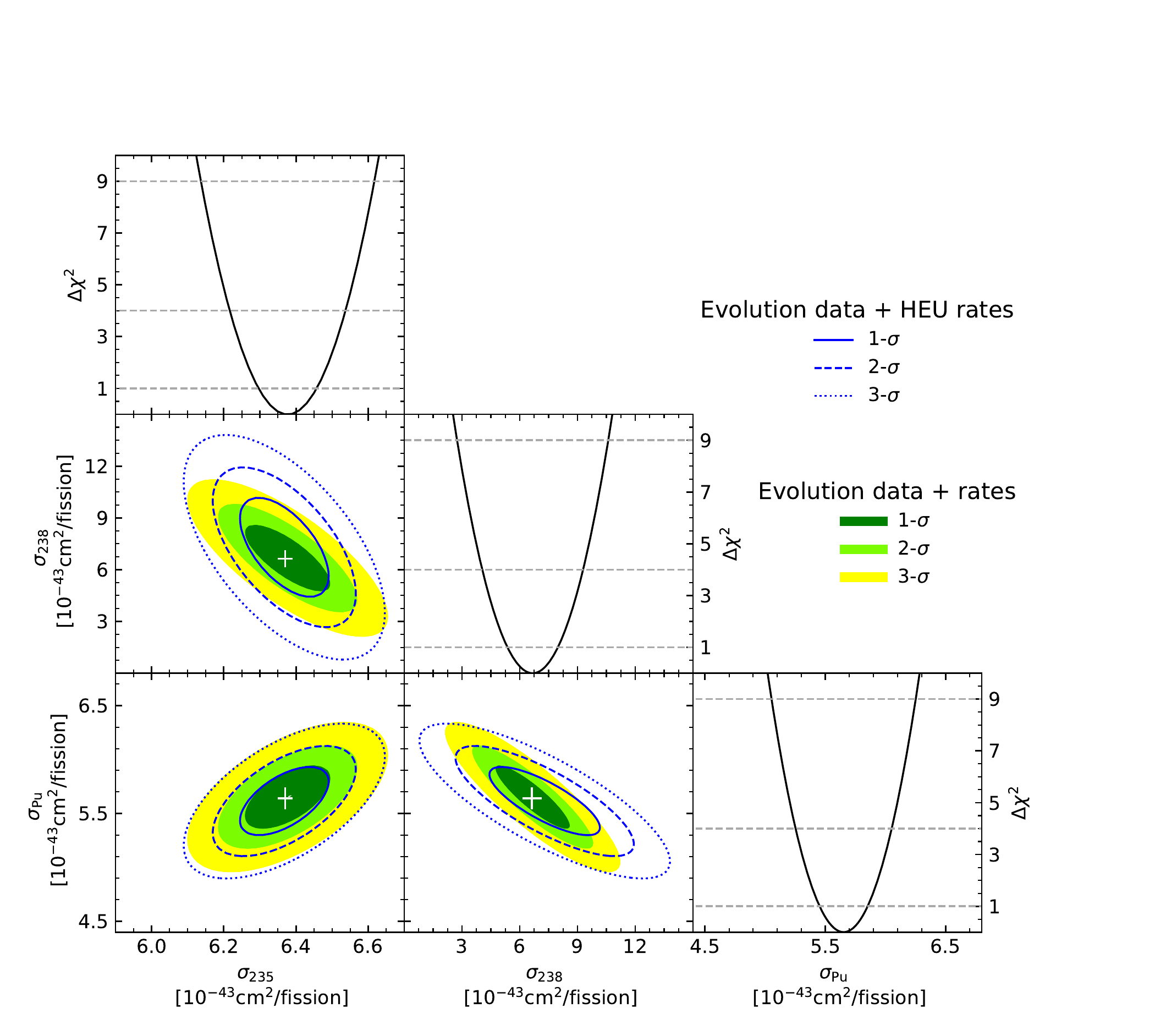}
    \caption{One-dimensional and two-dimensional marginalized distribution with $1\sigma$, $2\sigma$ and $3\sigma$ contours of $\sigma_{235}$, $\sigma_{238}$ and $\sigmacomb$ provided by a fit of \textbf{evolution data + reactor rates}. The blue solid, dashed and dotted contours show the allowed regions of \textbf{evolution data + HEU reactor rates} at $1\sigma$, $2\sigma$ and $3\sigma$ The darkgreen, lawngreen and yellow shaded regions show the results of the allowed regions of \textbf{evolution data + reactor rates}. The three black curves show one-dimensional $\Delta \chi^2$ profiles for $\sigma_{235}$, $\sigma_{238}$ and $\sigma_{\rm{Pu}}$ fitted with \textbf{evolution data + reactor rates}.}
    \label{fig.DYB_RENO_rate}
\end{figure*}

In this section, we shall present the analysis from the fits of reactor data including reactor rates listed in Table~\ref{tab.rates_2020} and reactor fuel evolution data from Daya Bay~\cite{DayaBay:2017jkb} and RENO~\cite{RENO:2018pwo}. 
Eq.~(\ref{eq.chi2_3d}) and Eq.~(\ref{eq.sigma_fa_k}) will be used to fit with the reactor data, and the HM model will be used for the corrections of residual $\Pum$. The fitting results are shown in Table~\ref{tab.results}, Figure~\ref{fig.DYB_RENO_HEU} and Figure~\ref{fig.DYB_RENO_rate}, and detailed analyses are provided in the following subsections.

In Section~\ref{subsec.HEU}, we show the fit of HEU reactor rates, which exclusively constrain the $\U$ isotopic IBD yield. In Section~\ref{subsec.evo_fit},
the combined fit of fuel evolution data and HEU reactor rates will be presented, which clearly illustrates the complementary role of each data set.
In Section~\ref{subsec.evo_rates}, the LEU reactor rate data will be added in the fit, providing a better constraint on the $\Um$ isotopic IBD yield.
The fitting results based on all data sets are also illustrated in Section~\ref{subsec.evo_rates}. Finally, in Section~\ref{subsec.app} we prove the robustness of the data-driven model against the model variations of the residual $\Pum$, and apply to the prediction of future reactor experiments.

\begin{table}[ht!]
    \centering
    \begin{tabular}{|c|c|c|c|c|c|}
    \hline
    \hline
         {} & HEU rates & Evolution data & Evolution data + HEU rates & LEU rates & Evolution data + rates \\
         \hline
         $\chi^2_{\text{min}}$ & \quad $8.9$ \quad & \quad $7.6$ \quad & \quad $17.5$ \quad & \quad $8.6$ \quad &\quad $25.7$ \quad \\
         NDF & \quad $7$ \quad & \quad $13$ \quad & \quad $21$ \quad & \quad $15$ \quad & \quad $37$ \quad \\
         GoF & \quad $26\%$ \quad & \quad $87\%$ \quad & \quad $68\%$ \quad & \quad  $90\%$ \quad & \quad $92\%$ \quad \\
         $\sigma_{235}$ & \quad $6.36\pm0.08$ \quad & \quad $3.60^{+2.7}_{-1.1}$ \quad & \quad $6.36\pm0.08$ \quad & \quad $6.85\pm0.55$ \quad  & \quad $6.37\pm0.08$ \quad \\
         $\sigma_{238}$ & \quad $\slash$ \quad & \quad $44.74^{+15}_{-37}$ \quad & \quad $7.25\pm1.88 $ \quad & \quad $0.68\pm8.81$ \quad  & \quad $6.63\pm1.30$ \quad \\
         $\sigma_{\rm{Pu}}$ & \quad $\slash$ \quad & \quad $1.52^{+4.0}_{-1.5}$ \quad & \quad $5.60\pm0.21$ \quad & \quad $6.37\pm2.13$ \quad  &\quad $5.64\pm0.20$ \quad \\
    \hline
    \hline
    \end{tabular}
    \caption{The model-independent IBD yields from the fits of different groups of data sets. The units are $\ibdunit$.}
    \label{tab.results}
\end{table}

\subsection{HEU reactors rates}
\label{subsec.HEU}

For HEU reactors (i.e., the research reactors), the antineutrinos are practically emitted from the beta decay generated from the fission products of $\U$. Thus in Eq.~(\ref{eq.chi2_3d}), we shall only work with the isotopic IBD yield of $\U$ and the theoretical prediction for HEU reactor rates in Eq.~(\ref{eq.sigma_fa_k}) can further be expressed as:
\begin{align}
    \sigma_{f,a}^{\text{fit}}= \sigma_{235}^{\text{fit}} \,. 
    \label{eq.sigma_fa_k_HEU}
\end{align}
The fitting results of HEU reactor rates are shown in the first column of Table~\ref{tab.results} and red lines in Figure~\ref{fig.DYB_RENO_HEU}. The red solid, dashed and dotted lines are illustrated as the allowed regions of the $\U$ isotopic IBD yield at 1$\sigma$, 2$\sigma$ and 3$\sigma$ confidence levels respectively. The best-fit value of $\sigma_{235}$ is $( 6.36\pm0.08 ) \, \ibdunit$,
as shown in the first column of Table~\ref{tab.results}, 
which is a deficit about $6\%$ compared to the prediction from the HM model, corresponding to a deviation about $4.7\, \sigma$ ($2.0\, \sigma$) lower than the HM model, where the first number is the deviation from the central value of the HM model, and the second number quoted in parentheses is the deviation from HM model considering the model uncertainty.
This result supports the hypothesis that $\U$ is the main isotope which is responsible to the reactor rate anomaly~\cite{Kopeikin:2021ugh,Giunti:2016elf,Giunti:2017nww,DayaBay:2017jkb,RENO:2018pwo}.

\subsection{Evolution data + HEU reactor rates}
\label{subsec.evo_fit}

The evolution of the experimental IBD yield during the reactor fuel burning in multiple burnup cycles has been measured at Daya Bay~\cite{DayaBay:2017jkb} and RENO~\cite{RENO:2018pwo}, which provides a powerful probe to constrain the reactor flux models and new physics beyond the SM~\cite{Giunti:2016elf,Giunti:2017nww,Hayes:2017res,Giunti:2017yid,Gebre:2017vmm,Giunti:2019qlt,Berryman:2020agd}. By constraining the isotopic IBD yields of $\Pum$ and $\Um$ from the HM model, it is shown that $\sigma_{235}$ is 8\% smaller than the HM model while $\sigma_{239}$ is consistent with the prediction~\cite{DayaBay:2017jkb,RENO:2018pwo}.
In order to remove the model dependence, we analyze the fuel evolution data from Daya Bay and RENO using three independent isotopic IBD yields $\sigma_{235}$, $\sigma_{238}$ and $\sigma_{\text{Pu}}$ defined in Eq.~(\ref{eq.sigma_fa_k}). The fitting results are shown in the second column of Table~\ref{tab.results} and blue lines of Figure~\ref{fig.DYB_RENO_HEU}, where the blue solid, dashed and dotted contours are the allowed regions of fuel evolution data at $1\sigma$, $2\sigma$ and $3\sigma$ confidence levels respectively. One can  observe that the fuel evolution data themselves cannot measure the individual isotopic IBD yields, but provide stringent constraints on the linear combinations of the three components. Therefore, in order to break the parameter degeneracy, it is quite important to combine the fuel evolution data and HEU reactor rates in the analysis.

In the third column of Table~\ref{tab.results} and shaded regions of Figure~\ref{fig.DYB_RENO_HEU}, we illustrate the combined analysis of the fuel evolution data and HEU reactor rates, where 
the darkgreen, lawngreen and yellow regions are shown for the $1\sigma$, $2\sigma$ and $3\sigma$ confidence levels respectively.
As expected, we can obtain the closed allowed regions for all three isotopic IBD yields. Firstly $\sigma_{235}= ( 6.36\pm0.08) \, \ibdunit$ is obtained, which is identical to that from the HEU rates-only analysis. Secondly, a meaningful measurement of $\sigma_{\text{Pu}}= (5.60\pm0.21) \, \ibdunit$ and $\sigma_{238}= (7.25\pm1.88) \, \ibdunit$ has been obtained by virtue of the synergy of two different data sets, in which the IBD yield of Pu's is 2.2\% larger than the HM prediction, and that of $\Um$ is less than the model by around 30\%, corresponding to the significance of $0.4\sigma$ and $1.4\sigma$ respectively, by taking account of the theoretical uncertainty of the HM model. 
Our results on $\sigma_{235}$ and $\sigma_{\text{Pu}}$ are both larger than those obtained in Ref.~\cite{DayaBay:2017jkb,RENO:2018pwo} by constraining the IBD yields of $\Pum$ and $\Um$, but their relative ratio is consistent with the evolution-only fit, which demonstrates that only the relative ratios of $\sigma_{235}$, $\sigma_{\text{Pu}}$ and $\sigma_{238}$ are properly constrained. With the inclusion of HEU reactor rates, the isotopic IBD yield of $\U$ is fixed and those of Pu's and $\Um$ are also decoupled, resulting in a complete and accurate determination of all three isotopic IBD yields. Note that the value of $\sigma_{238}$ deviates from the HM prediction with a significance of $1.4\sigma$. Whether this value is due to statistical fluctuation or hints at the problem of the $\Um$ flux requires future accurate measurements, in particular for those from HEU reactor rates.  

\subsection{Evolution data + reactor rates}
\label{subsec.evo_rates}

\begin{table*}[ht]
    \centering
    \begin{tabular}{|c|ccccc|}
    \hline
    \hline
    \bf Model & $\sigma_{235}$ & $\sigma_{238}$ & $\sigma_{239}$ & $\sigma_{241}$ &  $\sigma_{\text{Pu}}$\\
    \hline
    \bf This work & $6.37 \pm 0.08$ & $6.63 \pm 1.30$ & $\slash$ & $\slash$& $5.64\pm0.20$ \\
    \hline
    \bf HM & $6.74 \pm 0.17$ & $10.19 \pm 0.83$ & $4.40 \pm 0.13$ & $6.10 \pm 0.16$ & $5.48\pm0.16$\\
    \hline
    \bf EF & $6.29 \pm 0.31$ & $10.16 \pm 1.02$ & $4.42 \pm 0.22$ & $6.23 \pm 0.31$ & $5.52\pm0.23$\\
    \hline
    \bf HKSS & $6.82 \pm 0.18$ & $10.28 \pm 0.84$ & $4.45 \pm 0.13$ & $6.17\pm 0.16$ & $5.54\pm0.16$\\
    \hline
    \bf KI & $6.41 \pm 0.14$ & $9.53\pm0.48$ & $4.40 \pm 0.13$ & $6.10 \pm 0.16$ & $5.48 \pm 0.16$ \\
    \hline
    \hline
    \end{tabular}
    \caption{Comparison of this data-driven isotopic IBD yield model with the theoretical reactor flux models. The theoretical predictions of the HM, EF, HKSS, and KI models are taken from Ref.~\cite{Giunti:2021kab}. The IBD yields are given in units of $\ibdunits$.}
    \label{tab.model}
\end{table*}

In addition to the HEU reactor rates and fuel evolution data, there are also reactor rates data from LEU reactors with different values of fission fractions, as shown in Table~\ref{tab.rates_2020}. In the fourth column of Table~\ref{tab.results}, we present the fit of all LEU reactors, in which $\sigma_{235}$, $\sigma_{\text{Pu}}$ are nicely constrained, but $\sigma_{238}$ turns out to be vanishing and only upper bound is obtained due to the positivity of all the isotopic IBD yields.
In order to further reveal the role of LEU rates data,
the combined fit of fuel evolution data and HEU reactor rates is compared to the combined fit of fuel evolution data and all reactor rates in Figure~\ref{fig.DYB_RENO_rate}, where 
the blue solid, dashed and dotted contours show the allowed regions of the first combined fit at $1\sigma$, $2\sigma$ and $3\sigma$ confidence levels respectively, and the darkgreen, lawngreen and yellow shaded regions show the allowed regions of the second combined fit. Note that in the combined fit of evolution data and all reactor rates, two data points from the Daya Bay and RENO rates in Table~\ref{tab.rates_2020} have been excluded to avoid double counting.
From the figure one can observe that the inclusion of the LEU rates has negligible effects on the isotopic IBD yields of $\sigma_{235}$, $\sigma_{\text{Pu}}$, but the improvement on $\sigma_{238}$ is significant, where the relative uncertainty is reduced from 26\% to 19\%, and the significance of deviation from the HM model is also increased from $1.4\sigma$ to $2.3\sigma$, raising a question on the theoretical IBD yield of $\Um$ from both the summation and conversion models.
The best fits and associated uncertainties of $\sigma_{235}$, $\sigma_{\text{Pu}}$ and $\sigma_{238}$ from all reactor data are also provided in the last column of Table~\ref{tab.results}, where the isotopic IBD yields for 
$\U$, Pu's and $\Um$ are 
$\sigma_{235}=(6.37\pm0.08)\, \ibdunit$,
$\sigma_{\text{Pu}}=(5.64\pm0.20)\, \ibdunit$ and 
$\sigma_{238}=(6.63\pm1.30)\, \ibdunit$, respectively.
A comparison of this data-driven model with the theoretical reactor flux models is illustrated in Table~\ref{tab.model}, where the theoretical predictions of the HM, EF, HKSS, and KI models are taken from Ref.~\cite{Giunti:2021kab}.
From the table, it is shown that the data-driven IBD yield for $\U$ is consistent with EF and KI models within $1\sigma$, but smaller than HM and HKSS models by around 2-3$\sigma$. Meanwhile, the data-driven IBD yield for $\Um$ is also smaller than all the theoretical models by around 2-3$\sigma$, and $\sigma_{\text{Pu}}$ for Pu's is consistent with all models within $1\sigma$. These conclusions are consistent with those in Ref.~\cite{Giunti:2021kab}, but from a rather different viewpoint by decomposing the possible isotopic contributions.
Finally, from Figure~\ref{fig.DYB_RENO_rate} it is evident that possible correlations between different isotopic contributions should be considered when applying this data-driven model to the prediction of future reactor experiments.

\subsection{Discussion and application}
\label{subsec.app}

 \begin{table}
     \centering
     \begin{tabular}{|c|c|c|c|}
     \hline
     \hline
     {} & $\sigma^{\rm HM}_{241}$ ($\sigma^{\rm KI}_{241}$) & $\sigma^{\rm HKSS}_{241}$ & $\sigma^{\rm EF}_{241}$\\
     \hline
     $\chi^2_{\text{min}}$ & \quad $25.7$ \quad & \quad $25.7$ \quad & \quad $25.7$ \quad\\
     NDF & \quad $37$ \quad & \quad $37$ \quad & \quad $37$ \quad\\
     GoF & \quad $92\%$ \quad & \quad $92\%$ \quad  &\quad $92\%$ \quad\\
    $\sigma_{235}$ & \quad $6.37\pm0.08$ \quad & \quad $6.37\pm0.08$ \quad & \quad $6.37 \pm 0.08$ \quad\\
    $\sigma_{238}$ & \quad $6.63\pm1.30$ \quad & \quad $6.66 \pm 1.30$ \quad &\quad $6.68 \pm 1.30$ \quad \\
     $\sigma_{\rm{Pu}}$ &\quad $5.64\pm0.20$ \quad & \quad $5.63\pm0.20$ \quad & \quad $5.63\pm0.20$ \quad\\
     \hline
     \hline
     \end{tabular}
     \caption{
     The data-driven isotopic IBD yields from global fits of all reactor data based on the residual $\Pum$ corrections from different theoretical models in Ref.~\cite{Giunti:2021kab}. The IBD yields are given in units of $\ibdunits$.}
    \label{tab.results_from_other_models}
 \end{table}
As previously mentioned, in the analysis of the data-driven isotopic IBD yields, there is still a tiny level of model dependence on the corrections of the residual $\Pum$.
To prove the robustness of the data-driven model against the model variations, we have performed parallel global fits of all reactor data by using the residual $\Pum$ corrections from different theoretical models in Ref.~\cite{Giunti:2021kab}. The results are shown in Table~\ref{tab.results_from_other_models}, in which the variations of the data-driven isotopic IBD yields are as small as
$0.2\,\%$ and $0.7\,\%$ for $\sigma_{\rm{Pu}}$ and $\sigma_{238}$ respectively. Meanwhile $\sigma_{235}$ keeps unchanged as expected. In this case the variations for the expected IBD yields would be at the level of 0.1\% considering the typical fission fractions of commercial reactors. Therefore, the data-driven isotopic IBD yields are practically model-independent. In summary, we have demonstrated the robustness of the isotopic IBD yields for $\U$, $\Um$ and Pu's and are going to propose the following data-driven model for the isotopic IBD yields
\begin{align}
\left\{
\begin{aligned}
    \sigma_{235} & = ( 6.37 \pm 0.08 ) \, \ibdunit \\
    \sigma_{238} & = ( 6.63 \pm 1.30 ) \, \ibdunit \\
    \sigma_{\text{Pu}} & = ( 5.64 \pm 0.20 ) \, \ibdunit \,,
\end{aligned}
\right.
\label{eq.normimal}
\end{align}
and the corresponding correlation matrix
\begin{align}
    \rho = 
    \left(                
  \begin{array}{ccc}   
    1.000 & -0.751 & 0.571\\  
    -0.751 & 1.000 & -0.875\\  
    0.571 & -0.875 & 1.000\\ 
  \end{array}
\right)\,,
 \label{eq.correlation}
\end{align}
where $\sigma_i$'s ($i=235,\,238,\,\text{and} \, \text{Pu}$) are from the fits of all reactor data (i.e., evolution data + reactor rates) based on the residual $\Pum$ correction from $\sigma_{241}^{\text{HM}}$.

\begin{figure}[ht]
    \centering
    \includegraphics[width=0.6\textwidth]{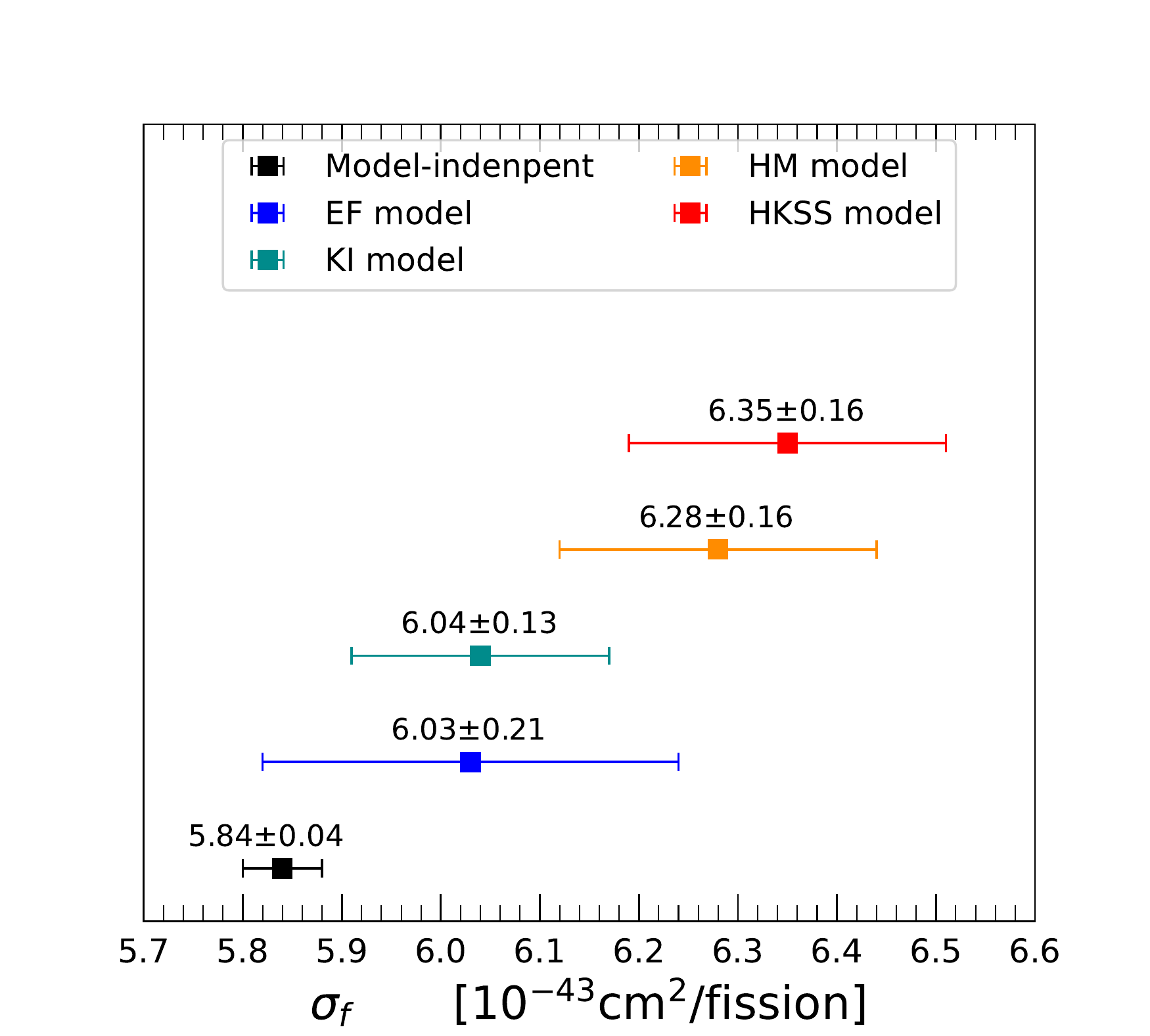}
    \caption{Predicted IBD yields for a future reactor experiment with both the model independent IBD yields and theoretical flux models in Ref.~\cite{Giunti:2021kab}.
    The IBD yields are given in units of $\ibdunits$.}
    \label{fig.app}
\end{figure}

Next as an application, we adopt the model-independent isotopic IBD yields to predict the expected IBD yield of a future reactor antineutrino experiment "$A$" with typical fission fractions ($f_{235}^A, \, f_{238}^A, \, f_{239}^A, \, f_{241}^A$):
\begin{align}
    \sigma_A = f_{235}^A \cdot \sigma_{235} + f_{238}^A \cdot \sigma_{238} + f_{239}^A \cdot \sigma_{\text{Pu}} + \Delta f^A \cdot \sigma_{241}^{\rm{HM}}\,.
\end{align}
By using the isotopic IBD yields from Eq.~(\ref{eq.normimal}) and the correlation matrix of Eq.~(\ref{eq.correlation}), we can estimate the predicted IBD yields of a future experiment with ($\U: \Um: \Pu : \Pum) = (0.577 : 0.076 : 0.295 : 0.052$)~\cite{JUNO:2015zny}:
\begin{align}
    \sigma_f = [5.84 \pm (0.04)_{\rm{MI}}\pm (0.0004)_{\rm{HM}}] \ibdunit \qquad (0.7\%\;{\rm precision})\,,
    \label{eq.pwr}    
\end{align}
where  the first and second terms represent the uncertainties originating from model-independent (MI) isotopic IBD yields and the HM model input of $\sigma_{241}$, respectively. Note that the latter one are completely negligible, and here a sub-percent precision (i.e., 0.7\%\;{\rm precision}) of the expected IBD yield has been obtained. A comparison of the predicted IBD yields using both the model-independent method and theoretical flux models in Ref.~\cite{Giunti:2021kab} is illustrated in Figure~\ref{fig.app}. From this comparison, we conclude that the model-independent method has the best precision among all the models, which is because the data-driven isotopic IBD yields have a specific form of the correlation matrix as in Eq.~(\ref{eq.correlation}), resulting in sizable cancellations between different isotopic contributions. 
Note that the model-independent method predicts a smaller IBD yield than the HM and HKSS models, free from the reactor antineutrino anomaly.

If the considered reactors are different from the pressurized water reactor (PWR), the linear relationship of Pu's cannot fully account for all the Pu contributions. Taking the European reactors with the mixed oxide (MOX) technology as an example, the typical fission fractions are given as $\U: \Um: \Pu: \Pum = 0.000: 0.080 : 0.708 : 0.212$~\cite{Borexino:2010dli}. In this case, the predicted IBD yield is
\begin{align}
    \sigma_f = \left[5.05\pm(0.07)_{\rm{MI}}\pm(0.01)_{\rm{HM}}\right] \ibdunit \qquad (1.4\%\, \text{precision})\,,
\label{eq.mox}    
\end{align}
which has an uncertainty of 1.4\% and the contribution from $\sigma^{\rm{HM}}_{241}$ is larger than that in Eq.~(\ref{eq.pwr}), but still much smaller than the model-independent part.

Finally before finishing this section, we would like to illustrate the robustness of the analysis framework used in this study, in which the model-independent isotopic IBD yields are obtained based on the linear relationship of Pu's and a fixed average ratio of $k$.
Firstly, we consider different values of $k$ in the linear fit of different data sets (i.e., $k=0.173, 0.177, 0.185$).
The model-independent isotopic IBD yields $\sigma_{235}$, $\sigma_{238}$ and the ratio of $r_{\text{Pu}}=\sigma_{\rm{Pu}}/(\sigma_{239}^{\rm{HM}}+k\cdot \sigma_{241}^{\rm{HM}} )$
will be essentially unchanged,
which demonstrates that the predicted IBD yield for a certain reactor neutrino experiment also keeps the same. Secondly, when a $6\%$ fractional uncertainty of $k$ is taken into account in the $\chi^2$ function with a pull term, the fitting results are practically the same as Table~\ref{tab.results} and Eq.~(\ref{eq.correlation}). Thirdly, a quadratic relationship of the fission fractions of Pu's is also considered, and the resulting IBD yield of a reactor neutrino experiment is also consistent with the cases of the linear relationship.
All these aforementioned stress tests demonstrate the robustness and credibility of the fitting framework, and ensure the model-independent isotopic IBD yields reliable.

\section{Conclusion}
\label{sec.conclusion}

The latest measurements from reactor antineutrino rates and spectra have shown anomalous results compared with the theoretical reactor flux models.  
In this work, we have obtained a model-independent reactor isotopic cross sections per fission from the global fits of the reactor antineutrino data from HEU reactor rates, LEU reactor rates, and reactor fuel evolution data. 
By using the implicit quasi-linear relationship between the fission fractions of $\Pu$ and $\Pum$ in the LEU reactor data, the IBD yields of $\U$, $\Um$, and Pu's can be properly constrained in the global fits, where the HEU reactor rates practically provide a complete determination of the isotopic IBD yield of $\U$, and the fuel evolution data put stringent constraints on the linear combinations of $\sigma_{235}$, $\sigma_{\text{Pu}}$ and $\sigma_{238}$. In addition, the inclusion of LEU reactor rates can further reduce the uncertainty of the isotopic IBD yield of $\Um$. We have also demonstrated the robustness of the data-driven isotopic IBD yields and proposed a model-independent method to calculate the expected IBD yield of future reactor experiments,
in which better than 1\% accuracy can be achieved for those using typical commercial reactors. Our method can also be generalized to the study of reactor antineutrino spectra from different reactor experiments or experiments with different fission fractions. It would be important to disentangle the isotopic contributions~\cite{Huber:2016xis,DayaBay:2019yxq} to the reactor shape anomaly, which will be reported in a future separated work.



\begin{acknowledgments}
Z.~Xin is grateful to Dr.~Jianrun Hu for the helpful discussion.
The work of Y.F.~Li and Z.~Xin was supported by National Natural Science Foundation of China under Grant Nos.~12075255 and 11835013, by Beijing Natural Science Foundation under Grant No.~1192019, by the Key Research Program of the Chinese Academy of Sciences under Grant No.~XDPB15.
Y.F. Li is also grateful for the
support by the CAS Center for Excellence in Particle Physics (CCEPP).
\end{acknowledgments}


\bibliographystyle{h-physrev5}
\bibliography{main}

\end{document}